\documentclass[%
reprint,
showpacs,%
showkeys,
amsmath,amssymb,
pre,
floatfix,
]{revtex4-1}

\usepackage[utf8]{inputenc}
\usepackage[T1]{fontenc}
\usepackage[english]{babel}
\usepackage{amsfonts}
\usepackage{graphicx}

\newcommand{\onefig}[1]{\centering{
    \includegraphics[width=0.96\columnwidth]{#1}}}

\newcommand{\supT}{\text{T}}
\newcommand{\partialt}[1]{\partial_t #1}

\newcommand{\partialxx}[1]{\partial^2_x #1}

\newcommand{\myii}{\mathrm{i}}
\newcommand{\mycolon}{\,\text{:}\,}
\newcommand{\mat}[1]{\mbox{\boldmath{$\mathrm{#1}$}}}
\renewcommand{\vec}[1]{\mbox{\boldmath{$#1$}}}
\newcommand{\bkwvec}{\vec{\varphi}^-}
\newcommand{\fwdvec}{\vec{\varphi}^+}

\newcommand{\bkwmat}{\mat{\Phi}^-} 
\newcommand{\fwdmat}{\mat{\Phi}^+}
\newcommand{\clvecfwd}{\vec{\gamma}}
\newcommand{\clmatfwd}{\mat{\Gamma}}
\newcommand{\determ}{\mathop{\text{det}}}
\newcommand{\subExpan}{1}
\newcommand{\subContr}{2}
\newcommand{\tqr}{T_\text{QR}}

\begin{document}

\title{Fast numerical test of hyperbolic chaos}

\author{Pavel V. Kuptsov}\email[Electronic
address:]{p.kuptsov@rambler.ru}%
\affiliation{Department of Instrumentation Engineering, Saratov State
  Technical University, Politekhnicheskaya 77, Saratov 410054,
  Russia}%




\date{\today}

\begin{abstract}
  The effective numerical method is developed performing the test of
  the hyperbolicity of chaotic dynamics. The method employs ideas of
  algorithms for covariant Lyapunov vectors but avoids their explicit
  computation. The outcome is a distribution of a characteristic value
  which is bounded within the unit interval and whose zero indicate
  the presence of tangency between expanding and contracting
  subspaces. To perform the test one needs to solve several copies of
  equations for infinitesimal perturbations whose amount is equal to
  the sum of numbers of positive and zero Lyapunov exponents. Since
  for high-dimensional system this amount is normally much less then
  the full phase space dimension, this method provide the fast and
  memory saving way for numerical hyperbolicity test of such systems.
\end{abstract}

\maketitle

\emph{Introduction. -- } The dynamics is said to be uniformly
hyperbolic if exponential rates of tangent space growth and
contractions are always bounded and differ from zero by some global
constants. Systems with hyperbolic dynamics admit rigorous proof of
their chaotic properties. The hyperbolic chaos is structurally stable,
i.e., it persists under change of parameters of a system.  Over many
years this paradigm remained mainly the theoretical, since the most of
the known systems do not conform with the basic assumptions of uniform
hyperbolicity. But recently the interest to hyperbolic chaos has been
renewed after a series of publications by Kuznetsov and his
collaborators who suggested sufficiently simple ideas of practical
implementation of the uniform hyperbolic chaos in natural
systems~\cite{KuzBook, KuzArt}.

Tangent space at each point of a hyperbolic chaotic trajectory is
split into invariant subspaces with different expanding and
contracting properties. For discrete time systems there are two
subspaces. The first one contains all expanding directions associated
with positive Lyapunov exponents, and the second one consists of the
all contracting directions corresponding to negative exponents. The
hyperbolicity implies the strict separation of these
subspaces~\cite{GuckHolm}. It means that the smallest angle between
expanding and contracting vectors are globally bound from zero. For
flows one more invariant neutral subspace is associated with zero
Lyapunov exponents. To extend the definition of the hyperbolic
dynamics to this case one have to require the strict separation of
these three subspaces~\cite{Pesin}.

Rigorous mathematical verification of hyperbolicity is not always
possible. To test this property numerically one can find bases for
expanding, contracting (and neutral, if any) subspaces and compute the
smallest angles between vectors from these subspaces for different
points along trajectories. If the dynamics is non-hyperbolic, among
trajectories there are ones with tangencies, i.e., with zero angles
between these vectors. Performing numerical simulations one normally
can not hit such trajectory exactly. But one can expect that randomly
chosen trajectory will pass infinitely close to the trajectories with
tangencies. Thus for non-hyperbolic dynamics the distributions of
smallest angles between subspaces have to be infinitely close to the
origin, while for hyperbolic dynamics these distributions are well
detached from zero.

This approach was initially developed for low-dimensional dynamics
(see~\cite{KuzBook} for review). Its application for high-dimensional
dynamics became possible only recently after discovery of effective
algorithms for computation of covariant Lyapunov vectors
(CLVs)~\cite{GinCLV,WolfCLV}. CLVs are associated with Lyapunov
exponents and are covariant with the tangent flow, i.e., the $i$-th
covariant vector at time $t_1$ is mapped to the $i$-th covariant
vector at $t_2$. These vectors provide a natural bases for expanding,
contracting, and neutral subspaces.

In paper~\cite{HyperSapce08} the angles between expanding and
contracting subspaces spanned by CLVs were analysed to detect
hyperbolic and non-hyperbolic regimes of a spatially extended
system. Also the angles between subspaces spanned by CLVs were
employed in Refs.~\cite{EffDim,ModesSplit10} to identify possible
bases for inertial manifolds~\cite{HypDecoupl11}.

Unfortunately, analysis of angles between subspaces spanned by CLVs is
very resource-consuming. One needs to process a lot of angles for
sufficiently long trajectory to obtain a representative
statistics. Moreover, it is usually unclear which CLVs from the
negative end of the spectrum can be safely omitted without distortion
of the result, and the most reliable way is to compute them all.

The other quantitative numerical test of hyperbolicity is based on the
direct verification of the cone criterion~\cite{KuzBook, KuzArt}. This
method is well developed for low-dimensional system and provides
perhaps the most reliable result. But unfortunately it is unclear yet
how to transfer it to high-dimensional systems. Moreover even if this
can be done, most probably it will also be a resource-consuming.

The paper~\cite{CLV2011} reviews the theory of Lyapunov vectors and
the related numerical algorithms. Moreover an improved approach for
computing CLVs is suggested. In this letter the idea of this approach
is employed to develop the fast method for testing of the
hyperbolicity without explicit computing the CLVs. Given the system
with $k_+$ positive and $k_0$ zero Lyapunov exponents, it is enough to
solve only $k_++k_0$ equations for infinitesimal perturbations. Since
for high-dimensional systems $k_++k_0$ is normally much less then the
whole phase space dimension, this method provides the fast and memory
saving way for numerical hyperbolicity test of such systems.

\emph{Theory. -- } Preforming the standard procedure of computation of
Lyapunov exponents, we deal with a set of orthonormal vectors which
are evolved in tangent space by alternating time evolution over an
interval $\tqr$ and re-orthogonalization via Gram-Schmidt or QR
procedure. After sufficiently many iterations these vectors converge
to backward Lyapunov vectors (also they are called Gram-Schmidt
vectors). They are ``backward'' because initialized in the far past.
Let $\bkwvec_i(t)$ be the $i$-th backward Lyapunov vector and
$\bkwmat(t)=[\bkwvec_1(t),\bkwvec_2(t),\ldots,\bkwvec_m(t)]$ be an
orthogonal matrix of these vectors, where $m$ is the dimension of the
tangent space. Performing the algorithm for Lyapunov exponents
backward in time we obtain in the same way another set of orthonormal
vectors $\fwdvec_1(t)$ and the corresponding orthogonal matrix
$\fwdmat(t)=[\fwdvec_1(t),\fwdvec_2(t),\ldots,\fwdvec_m(t)]$.  These
vectors are referred to as forward Lyapunov vectors, because they are
initialized in the far future. Both $\bkwvec_i(t)$ and $\fwdvec_i(t)$
are associated with the $i$-th Lyapunov exponent $\lambda_i$. The
Lyapunov exponents are assumed to be ordered as usual in descending
order.

Let $\clmatfwd(t)=[\clvecfwd_1(t),\clvecfwd_2(t),\ldots,\clvecfwd_m]$
be a matrix of the CLVs $\clvecfwd_i(t)$. These vectors are related to
the forward and backward Lyapunov vectors as~\cite{WolfCLV,CLV2011}
\begin{equation}\label{eq:clv_defin}
  \clmatfwd(t)=\bkwmat(t)\mat{A}^-(t)=\fwdmat(t)\mat{A}^+(t),
\end{equation}
where $\mat{A}^-(t)$ is an upper triangular matrix and $\mat{A}^+(t)$
is a lower triangular matrix. In fact $\bkwmat(t)\mat{A}^-(t)$ is the
QR decomposition of $\clmatfwd(t)$ while $\fwdmat(t)\mat{A}^+(t)$ is
its QL decomposition.

Given the matrices $\bkwmat(t)$ and $\fwdmat(t)$ one can compute
covariant Lyapunov vectors using the equation
$\mat{P}(t)\mat{A}^-(t)=\mat{A}^+(t)$, where
\begin{equation}\label{eq:def_mat_p}
  \mat{P}(t)=[\fwdmat(t)]^\supT\bkwmat(t)
\end{equation}
is a $m\times m$ orthogonal matrix, and $\mat{A}^\pm(t)$ are
components of the LU decomposition of $\mat{P}(t)$. The details of
this LU-method of computation of CLVs can be found in~\cite{CLV2011}.

Let us assume that we have the whole set of $m$ CLVs. Consider two
subspaces of the tangent space. The first one is spanned by the first
$k$ CLVs, and the second one is spanned by the rest of them. We need
to check if there is a tangency between vectors from these subspaces.
Consider two arbitrary unit vectors from these subspaces:
\begin{equation}
  \vec v_\subExpan(t)=\sum_{i=1}^k e_i\clvecfwd_i(t), \;\;
  \vec v_\subContr(t)=\sum_{i=k+1}^m c_i\clvecfwd_i(t),
\end{equation}
where $e_i$ and $c_i$ are some expansion coefficients. To detect the
tangency we can compute the angle between $\vec v_\subExpan(t)$ and
$\vec v_\subContr(t)$ and find such coefficients $e_i$ and $c_i$ that
minimize is. Zero minimal angle indicates the tangency.

As follows from Eq.~\eqref{eq:clv_defin}, the covariant vector
$\clvecfwd_j(t)$ belongs to the subspace spanned by the first $j$
forward Lyapunov vectors $\bkwvec_i(t)$, where $i=1,2,\ldots,j$. Also
$\clvecfwd_n(t)$ belongs to the subspace of the rest $m-n+1$ forward
vectors $\fwdvec_i(t)$, where $i=n,n+1,\ldots,m$. It means that $\vec
v_\subExpan(t)$ can be represented as a linear combination of the
backward Lyapunov vectors and $\vec v_\subContr(t)$ is a linear
combination of the forward Lyapunov vectors,
\begin{equation}
  \vec v_\subExpan(t)=\sum_{i=1}^k e'_i\bkwvec_i(t), \;\; 
  \vec v_\subContr(t)=\sum_{i=k+1}^m c'_i\fwdvec_i(t),
\end{equation}
with some expansion coefficients $e'_i$ and $c'_i$.

The tangency occurs if there are such $e'_i$ and $c'_i$ that $\vec
v_\subExpan(t)=\vec v_\subContr(t)$:
\begin{equation}
  \label{eq:v1v2}
  \sum_{i=1}^k e'_i\bkwvec_i(t)=\sum_{i=k+1}^m c'_i\fwdvec_i(t).
\end{equation} 
To find $e'_i$ and $c'_i$ we multiply \eqref{eq:v1v2} by vectors
$\fwdvec_j(t)$. The first $k$ multiplications result in the set of
equations for $e'_i$ and the rest of them produces the equations for
$c'_i$, provided that the nontrivial solution for $e'_i$ exists.  The
equations for $e'_i$ read:
\begin{equation}
  \sum_{i=1}^k e'_i\fwdvec_j(t)\bkwvec_i(t)=0, \;\; j=1,2\ldots,k.
\end{equation}
The nontrivial solution exists, when the scalar products
$\fwdvec_j(t)\bkwvec_i(t)$, where $1\leq i,j\leq k$, form a singular
matrix. One can see from Eq.~\eqref{eq:def_mat_p}, that this matrix is
the top left $k\times k$ submatrix of $\mat P$. This provides an idea
of the test of tangencies between subspaces and, in particular, the
test of the hyperbolicity.

\emph{Formulation of the method. -- } Given a dynamical system, in its
tangent space we need to find the smallest angle between vectors from
the subspace spanned by the first $k$ covariant vectors and from the
subspace spanned by the rest $m-k$ of them. (a) We begin to move
forward in time solving simultaneously the basic equations and $k$
sets of linearized equations for infinitesimal perturbations. In the
same way as in computing Lyapunov exponents, we alternate time
evolution over intervals $\tqr$ and orthogonalizations via QR or
Gram-Schmidt algorithms. These steps are repeated for a while until
columns of orthogonalized matrices converge to $\bkwvec_i(t)$. (b) The
same procedure is continued but now $\bkwvec_i(t)$ are saved. These
steps are repeated as many times as many points of the trajectory we
are going to test. (c) We proceed to move forward along the trajectory
with the basic system only and save the trajectory points. The
equations for perturbations are not solved on this stage. (d) We turn
back and start to move backward along the saved trajectory performing
steps with the basic system and $k$ copies of equations for
perturbations. Again time evolution alternates with
orthogonalizations. In this way orthogonal vectors converge to
$\fwdvec_i(t)$. There two subtle points here. First, the modified
perturbation equations have to be used. The Jacobian matrix that
determines these equations have to be transposed and its elements have
to change their signs. In this case the vectors $\fwdvec_i(t)$ are
computed in the proper order, i.e., $k$ perturbation equations produce
\emph{the $k$ first vectors}, and not the last ones
(see~\cite{CLV2011} for explanations). Second, dealing with the
dissipative systems, it is better to use saved trajectory points
avoiding solving the basic system backward in time. This is because
negative Lyapunov exponents have typically large absolute values, and
this can result in the dramatic instability of the numerical
procedure. (e) After the arrival at the time step for which
$\bkwvec_i(t)$ was previously saved, we proceed the above procedure,
but additionally construct submatrices $\mat P(1\mycolon k,1\mycolon
k)$, see Eq.~\eqref{eq:def_mat_p}, and compute their normalized
determinants as
\begin{equation}
  \label{eq:nrm_det}
  d_k=|\determ[\mat P(1\mycolon k,1\mycolon k)]|/k!
\end{equation}
Here $0\leq d_k\leq 1$ since absolute values of elements of $\mat P$
are less or equal to 1. (f) We accumulate many $d_k$ and compute their
distribution. If this distribution is well separated from the origin,
the trajectory never pass close to points of tangencies between two
studied subspaces. If the amount of the processed points is
sufficiently large, this is the reason to conjecture that the
tangencies are absent at all. Otherwise, the close approach of the
distribution to the origin is the reason to conjecture that there are
trajectories with the exact tangencies.

Testing the hyperbolicity of chaos, we need to know the amount of
positive $k_+$ and negative $k_0$ Lyapunov exponents. If zero
exponents are absent, the distribution of $d_{k_+}$ have to be
analysed. In presence of zero exponents, there are two distributions
to consider: for $d_{k_+}$ and for $d_{k_{+}+k_0}$. For hyperbolic
chaotic systems both of the distributions are well separated from the
origin. Since the number of expanding and neutral directions are
normally much less then the full dimension of the tangent space, the
represented algorithm provide sufficiently fast way of testing the
hyperbolicity even for high-dimensional systems.

\begin{figure}
  \onefig{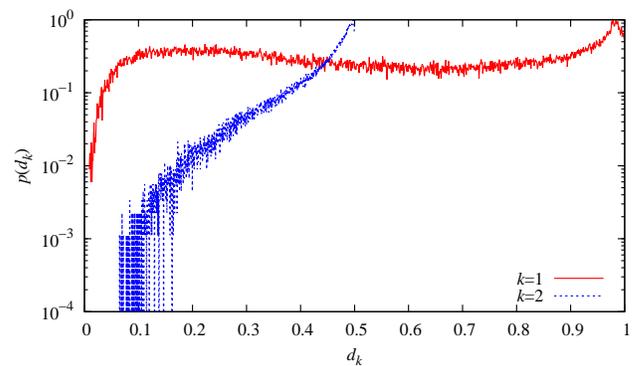}
  \caption{(color online). Distributions of normalized determinants
    \eqref{eq:nrm_det} over a trajectory of the Lorenz
    system~\eqref{eq:lor}. $\tqr=0.1$, $\Delta t\approx 0.002$.}
  \label{fig:lor}
\end{figure}

\emph{Examples. -- } First we consider the Lorenz system
\begin{equation}
  \label{eq:lor}
  \dot x=\sigma(y-x),
  \dot y=rx-xz-y,
  \dot z=xy-bz.
\end{equation}
where $r=28$, $b=8/3$, $\sigma=10$. The Lyapunov exponents are $0.90$,
$0$, and $-14.57$. There is one positive and one zero exponent and
thus we have to analyze distributions of $d_1$ and $d_2$, see
Fig.~\ref{fig:lor}. This system is known to be singular hyperbolic,
which means its tangent space admits an invariant splitting $E^c\oplus
E^{ne}$ into a 1-dimensional uniformly contracting sub-space and
2-dimensional volume-expanding subspace~\cite{SingHyp}. In an
agreement with this definition in Fig.~\ref{fig:lor} the distribution
$p(d_1)$ touches the origin, while $p(d_2)$ does not. It indicates the
existence of trajectories with tangencies between the expanding and
neutral subspaces, while the contracting subspace remains isolated.

\begin{figure}
  \onefig{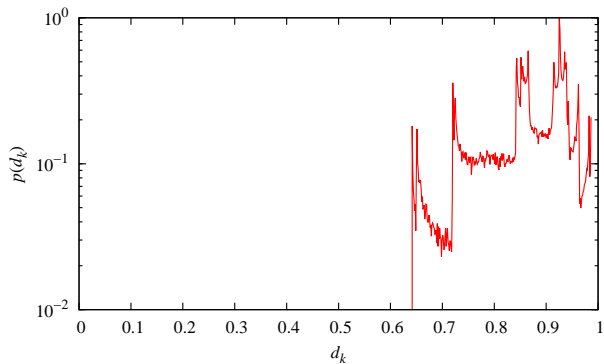}
  \caption{(color online). Distribution of $d_k$ for the stroboscopic
    map built for the system~\eqref{eq:kvdp} at $t=t_n=nT$. $\tqr=T$,
    $\Delta t\approx 0.01$.}
  \label{fig:kvdp}
\end{figure}

The other example is the system of two alternately excited van der Pol
oscillators
\begin{equation}
  \label{eq:kvdp}
  \begin{aligned}
    &\ddot x-[A\cos(2\pi t/T)-x^2]\dot x+\omega_0^2x=
    \epsilon y\cos\omega_0 t,\\
    &\ddot y-[-A\cos(2\pi t/T)-y^2]\dot y+4\omega_0^2y=\epsilon x^2.
  \end{aligned}
\end{equation}
where $A=5$, $T=6$, $\epsilon=0.5$, $\omega_0=2\pi$. Corresponding to
this system stroboscopic map at $t=t_n=nT$ is known to be
hyperbolic~\cite{KuzBook, KuzArt}. The Lyapunov exponents for this map
are $0.68$, $-2.61$, $-4.61$, $-6.10$. There is one positive exponent
and the zero one is absent since the system is non-autonomous. Hence
the hyperbolicity test for this map includes the analysis of
$p(d_1)$. As we see in Fig.~\ref{fig:kvdp} values of $d_1$ are located
far from the origin, confirming the applicability of the method.

\begin{figure}
  \onefig{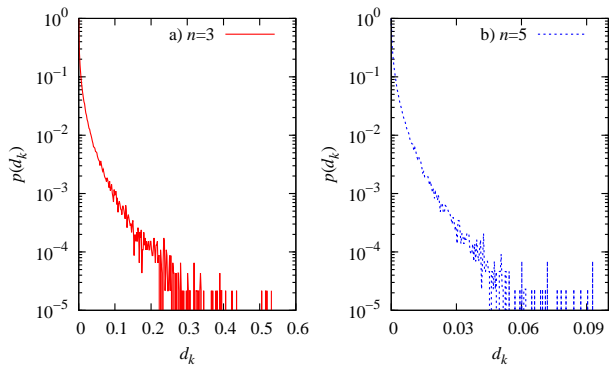}
  \caption{(color online). Distributions of (a) $d_3$ and (b) $d_5$
    for complex Ginzburg-Landau equation~\eqref{eq:cgle}. $\tqr=0.5$,
    $N=128$, $\Delta x=0.1$, $\Delta t\approx 0.0005$.}
  \label{fig:cgle}
\end{figure}

Next we consider the complex Ginzburg-Landau equation
\begin{equation}
  \label{eq:cgle}
  \partialt{a}=a-(1+\myii c)|a|^2 a+(1+\myii b)\partialxx{a}
\end{equation}
at $c=3$, $b=-2$, that corresponds to the amplitude
chaos~\cite{CgleChaos}. Since the no-flux boundary conditions are
used, the system has two continues symmetries, i.e., time translation
and phase rotation, and thus has two zero Lyapunov
exponents~\cite{SymCGLE1}. For chosen parameter values the first seven
exponents read: $0.30$, $0.18$, $0.060$, $0$, $0$, $-0.085$,
$-0.27$. It means that we have to consider $p(d_3)$ and $p(d_5)$, see
Fig.~\ref{fig:cgle}. For both distributions $d_k$ are essentially
nonzero at the origin, thus we have to conclude that the chaos is
essentially non-hyperbolic.

\begin{figure}
  \onefig{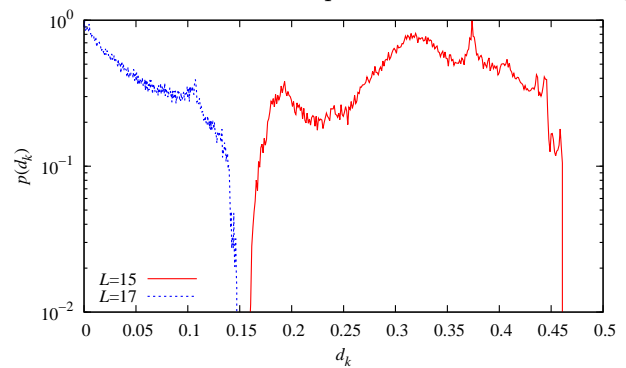}
  \caption{(color online). Distribution of $d_k$ for alternately
    excited Ginzburg-Landau equation~\eqref{eq:kuzcgle}. $\tqr=T$,
    $N=51$, $\Delta x=L/(N-1)$, $\Delta t\approx 0.05$.}
  \label{fig:kuzcgle}
\end{figure}

Finally, we apply the test to the alternately excited Ginzburg-Landau
equations, which are the amplitude equations for the
system~\eqref{eq:kvdp} supplied with diffusive terms:
\begin{equation}
  \label{eq:kuzcgle}
  \begin{aligned}
    &\partialt{a}=A\cos(2\pi t/T)a-|a|^2 a-\myii\epsilon b+\partialxx{a}, \\
    &\partialt{b}=-A\cos(2\pi t/T)b-|b|^2 b-\myii\epsilon
    a^2+\partialxx{b},
  \end{aligned}
\end{equation}
where $A=3$, $T=5$, $\epsilon=0.05$. The spatio-temporal chaos in this
system is hyperbolic when its length $L$ is sufficiently small and the
hyperbolicity is destroyed when the length
growths~\cite{HyperSapce08}. We consider the same parameters and
numerical mesh as in~\cite{HyperSapce08}. For $L=15$ first five
Lyapunov exponents, computed for the stroboscopic map at $t=t_n=nT$,
are $0.69$, $0.69$, $-0.11$, $-0.61$, $-1.62$. For $L=17$ the
exponents are $0.69$, $0.67$, $0.13$, $-0.28$, $-0.85$. Hence we have
to consider $p(d_2)$ for the first case and $p(d_3)$ for the second
case, see Fig.~\ref{fig:kuzcgle}. One can see that represented method
detects the reported in Ref.~\cite{HyperSapce08} transition from the
hyperbolic chaos at $L=15$ to the non-hyperbolic one at $L=17$.

Altogether we see that the demonstrated examples agree well with the
previously known results. The method works sufficiently fast. All
distributions represented above have been computed for $10^5$ values
of $d_k$ using 2.6~GHz processor. For distributed systems it took
approximately 3 hours.

The research is supported by RFBR-DFG grant 11-02-91334.

\end{document}